\documentclass{ws-procs9x6}

\usepackage{nicefrac}

\begin{document}

\title{The `PAMELA anomaly' indicates a nearby cosmic ray accelerator}

\author{P. MERTSCH$^*$ and S. SARKAR}

\address{Rudolf Peierls Centre for Theoretical Physics, University of Oxford,\\
Oxford OX1 3NP, UK\\
$^*$E-mail: p.mertsch1@physics.ox.ac.uk}

\begin{abstract}
  We discuss the recently observed `excesses' in cosmic ray electron
  and positron fluxes which have been widely interpreted as signals of
  dark matter. By considering the production and acceleration of
  secondary electrons and positrons in nearby supernova remnants, we
  predict an additional, harder component that becomes dominant at
  high energies. The unknown spatial distribution of the supernova
  remnants introduces a stochastic uncertainty which we estimate
  analytically. Fitting the prediction for different source
  distributions to the total electron + positron flux measured by
  Fermi--LAT fixes all free parameters and allows us to `postdict' the
  rise in the positron fraction seen by PAMELA. A similar rise in the
  B/C ratio is predicted at high energies.
\end{abstract}


\section{Introduction}

The recent findings of a rise in the positron fraction by
PAMELA~\cite{Adriani:2008zr} and an excess in the total $e^{+} +
e^{-}$ flux by Fermi--LAT~\cite{Abdo:2009zk} have generated a lot of
interest because these might be signals of dark matter annihilation or
decay. However the expected signals are much smaller than those
observed and models which attempt to circumvent this by invoking
exotic new physics are increasingly constrained by $\gamma$--ray and
radio observations. It is thus important to investigate more prosaic
astrophysical explanations, e.g. nearby
pulsars~\cite{Aharonian:1995zz,Hooper:2008kg} or supernova remnants
(SNRs). In this context we wish to revisit the predictions of the
standard SNR origin model of galactic cosmic rays (GCRs) wherein only
primary electrons and nuclei are accelerated in SNRs by diffusive
shock acceleration (DSA) \cite{Malkov:2001}.

We consider the recent proposal \cite{Blasi:2009hv} that some of the
secondary $e^{+}$ made \emph{in} the SNRs are also accelerated by
DSA. By accounting for the discrete distribution of the SNRs, we are
then able to match the predicted fluxes to the measured $e^{+} +
e^{-}$ flux. Having thus fixed all the free parameters of our model we
can make an independent `postdiction' for the positron fraction which
agrees remarkably well with the PAMELA data. We also predict a similar
rise for the B/C ratio as a definitive test of our model.

\section{The discreteness of sources}

Although SNRs can grow to become quite large (over $\sim 100 \,
\text{pc}$ in diameter) towards the end of their lifetime, they are
still much smaller than the Kpc scales relevant for GCR
propagation. Furthermore, the bulk of the particles accelerated by a
SNR will be released towards the end of the Sedov--Taylor phase when
they can no longer be confined by magnetic turbulence. We are
therefore not dealing with a continuous distribution of sources within
the galactic disk but with a large number of \emph{discrete} sources,
both in space and time. However, as we do not know the exact
distribution, this discreteness introduces an uncertainty in the
predicted total $e^{+} + e^{-}$ flux. However, a generalised central
limit theorem and the analytical form of the Green's function for the
diffusion--energy loss problem can be used to infer the statistical
properties of the flux, i.e. its average and quantiles. Our results
are in accordance with earlier studies using the Monte Carlo
approach~\cite{Pohl:1998ug,Strong:2001qp,Swordy:2003ds}.

The flux from a source $i$ that injected $e^{-}$ or $e^{+}$ a time
$t_i$ ago at a distance $L_i$ from the observer is given by the
Green's function~\cite{Ginzburg:1990sk} $G_\text{disk}(E,L_i,t_i)$ of
the diffusion equation: $J_i(E) = c/(4 \pi)
G_\text{disk}(E,L_i,t_i)$. The flux $J(E)$ of $N$ identical sources at
distances $\{ L_i \}$ and times $\{ t_i \}$ is just the sum of the
individual fluxes,
\begin{align}
\label{mer:SumOfFluxes}
J = \sum_{i=1}^N J_i(E) = \frac{c}{4 \pi} \sum_{i=1}^N
G_\text{disk}(E,L_i,t_i) \, .
\end{align}
As a function of the random variables $L$ and $t$, the Green's
function $Z \equiv G_\text{disk} (E, L, t)$ is itself a random
variable with probability density $f_Z$, expectation value $\mu_Z$ and
standard deviation $\sigma_Z$. If the central limit theorem were
applicable, the fluxes $J$ for different realisations of the same
source density at a fixed energy $E$ would follow a normal
distribution with mean $\mu_J = c/(4 \pi) N \mu_Z$ and standard
deviation $\sigma_J = c/(4 \pi) \sqrt{N} \sigma_Z$.

In the case under consideration, $L$ and $t$ are assumed to be
independent random variables with probability densities $f_L$ and
$f_t$, respectively. We choose the sources to be distributed
homogeneously in a ring with inner and outer radius $r_1$ and $r_2$,
so $f_L= 2 L / (r_2^2 - r_1^2)$ for $r_1 \leq r < r_2$ and $0$
otherwise. For simplicity we also assume the source rate to be
constant in time, i.e. $f_t = 1/t_{\mathrm{max}}$ for $0 \leq t \leq
t_{\mathrm{max}}$ and $0$ otherwise, up to a maximum time set by the
minimum energy $E_{\mathrm{min}}$ under consideration:
$t_{\mathrm{max}} = (b_0 E_{\mathrm{min}})^{-1}$. Here, $b_0$ is the
normalisation of the energy loss rate of the $e^-/e^+$: $b(E) = b_0
E^2$.

For $E > 10 \, \text{GeV}$ and $r_1 \rightarrow 0$ (the distance to
the nearest source is not limited physically) we find the expectation
value:
\begin{align}
  \mu_J = \frac{c}{4 \pi} \frac{1}{\sqrt{4 D_0 b_0 (1 - \delta)}}
  \frac{N}{\pi r_2^2} \frac{Q_0}{t_\text{max}} E^{- \gamma - 1 + (1 -
    \delta) / 2} \frac{\Gamma \left( \frac{\gamma - 1}{1 - \delta}
    \right) }{ \Gamma \left( \frac{2 \gamma - \delta - 1}{2 (1 -
        \delta)} \right) } \, .
\end{align}
where the interstellar diffusion coefficient $D_0 = 10^{28} \,
\text{cm}^2 \, \text{s}^{-1}$ at $E = 1 \, \text{GeV}$, and $\delta
\simeq 0.6$ is its spectral index ($D \propto E^\delta$). We set $b_0
= 10^{-16}\,\text{GeV}^{-1} \, \text{s}^{-1}$ and $r_2 = 15 \,
\text{kpc}$ as is standard. The total number $N$ of sources that are
needed to reproduce the (observed) number $\mathcal{N}~\simeq~300$ of
SNRs active in the Galaxy at any given time depends on the average
lifetime of a SNR, $\tau_\text{SNR}$, which is suggested to be $\sim
10{^4} \, \text{yr}$ \cite{Reynolds:2008}, hence $N = \mathcal{N}
t_{\mathrm{max}} / \tau_{\mathrm{SNR}} = 10^6$. With $Q_0 = 8.4 \times
10^{49}\,\text{GeV}^{-1}$, the expectation value,
\begin{align}
  E^3 \mu_J \simeq 150 \, \text{GeV}^{-1} \, \text{cm}^{-2} \,
  \text{s}^{-1} \, \text{sr}^{-1} \, ,
\end{align}
closely matches the featureless $E^{-3}$ spectrum measured by
Fermi--LAT~\cite{Abdo:2009zk}.

\begin{figure}
\begin{center}
\psfig{file=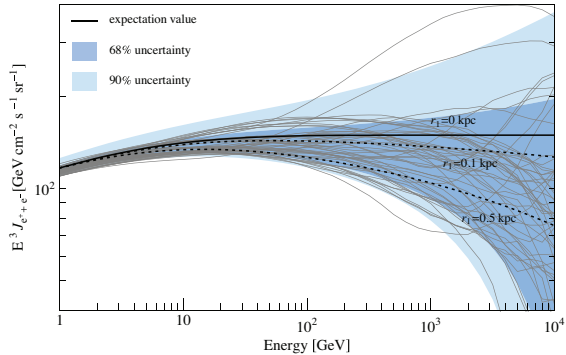,width=0.65\textwidth}
\end{center}
\caption{Fluxes of high energy $e^{-}$ from ensembles of sources
  uniformly distributed in a disk around the observer. The solid line
  denotes the expectation value for the sum of fluxes from $N$
  discrete, transient sources while the dashed lines show the
  expectation values if the sources are limited to a ring with inner
  radius $r_1$ (normalised to the expectation value for $r_1 = 0$ at
  $1 \, \text{GeV}$). The coloured bands quantify the fluctuations and
  contain, respectively, $68 \, \%$ and $90 \, \%$ of the calculated
  fluxes. The fluxes from 50 random realisations of an ensemble of $N$
  individual sources are shown by the thin grey lines.}
\label{mer:nuFluctuationsSpec}
\end{figure}

Whereas the expectation value $\mu_J$ is well-defined, the variance
$\sigma_J^2$ diverges because $f_Z(z)$ has a long power law
tail. Nevertheless, instead of the central limit theorem, one can
apply a generalised version \cite{Gnedenko:1954} which allows
determination of the distribution of $J$ from the asymptotic behaviour
of $f_Z(z)$ for $z \rightarrow \infty$. The distribution of $J$ is a
so-called stable distribution~\cite{Nolan:2010} --- an asymmetric
generalisation of a Gaussian which also exhibits a power law behaviour
for large values of $J$. The uncertainty interval around the
expectation value $\mu_J = c/(4 \pi) N \mu_Z$ can then be defined by
quantiles of the stable distribution. The energy dependence of the
quantiles turns out to be always harder than that of the expectation
value, so the fluctuations are growing with
energy. Figure~\ref{mer:nuFluctuationsSpec} shows the uncertainty bands
around the expectation value for the flux $\mu_J$, together with the
expected fluxes from 50 random realisations of the above source
distribution.

To calculate the fluxes of $e^{-}$ and $e^{+}$ at Earth, we perform a
Monte Carlo calculation by considering a large number of realisations
of randomly distributed sources according to a probability density
function which reflects our present knowledge concerning the
distribution of SNRs in the Galaxy. We emphasise that the better the
expected flux of $e^{-}$ and $e^{+}$ from such a realisation of the
source density matches the measured fluxes, the closer is the
underlying distribution of sources likely to be to the actual one. We
do not consider any variations between the SNRs but assume a
prototypical set of source parameters which we determine from a
compilation of $\gamma$-ray SNRs, see
Sec.~\ref{mer:PrimaryElectrons}. Of course all SNRs are not the same,
however variations of the source parameters would only introduce
additional fluctuations in the fluxes without altering their average.

\begin{figure}[tb]
\begin{tabular*}{\textwidth}{p{0.475\textwidth} p{0.475\textwidth}}
\psfig{file=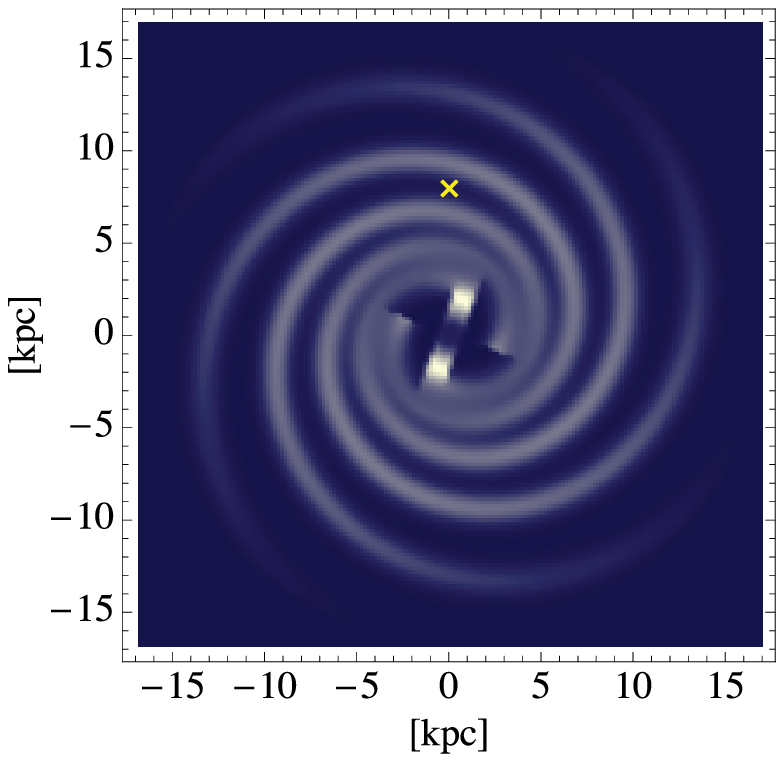,width=0.4\columnwidth}
&
\psfig{file=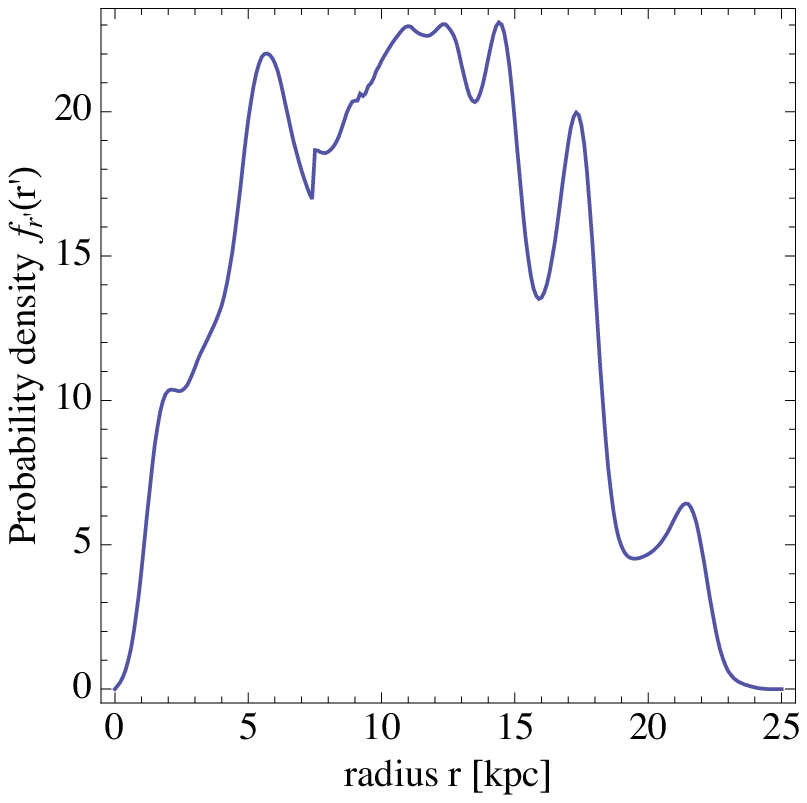,width=0.384\columnwidth}
\\
\begin{minipage}[t]{\linewidth}
\caption{The adopted distribution of SNRs in the Galaxy ---
  the cross denotes the position of the Sun.}
 \label{mer:spiral_structure}
\end{minipage}
&
\begin{minipage}[t]{\linewidth}
\caption{The probability density for the distance of a SNR
  from the Sun.}
\label{mer:SNR_density}
\end{minipage}
\end{tabular*}
\end{figure}

For a more realistic distribution of source distances, we model the
SNR density by a logarithmic spiral tracing the gas density (see
Fig.~\ref{mer:spiral_structure}) and also include the known radial
distribution of SNRs in the Galaxy from radio surveys. We transform to
a coordinate system centred on the Sun and integrate over azimuth such
that our calculated probability density (Fig.~\ref{mer:SNR_density})
encodes the average surface density of SNRs at a particular distance.

\section{Primary electrons}
\label{mer:PrimaryElectrons}

The injection of primary $e^{-}$ by SNRs is parametrised as:
\begin{equation}
  R_{e^-} = R_{e^-}^0 \left(\frac{E}{\text{GeV}}\right)^{-\gamma} 
   \text{e}^{-E/E_\text{cut}}.
\end{equation}
The spectral index $\gamma$ and the cut-off energy $E_{\mathrm{cut}}$
can be obtained from the spectral indices of SNRs in $\gamma$-rays as
measured by Imaging Air Cerenkov Telescopes (IACTs), like HESS, MAGIC
and VERITAS. (We assume the same spectral index for the hadronic and
the electronic components, as predicted for DSA.) We have
compiled~\cite{Ahlers:2009ae} a list of all SNRs detected by IACTs and
find the typical values to be $\gamma = 2.4$ and $E_{\mathrm{cut}} =
20 \, \text{TeV}$. The normalisation $R_{e^-}^0$ is determined by
fitting the electron flux at Earth from our Monte Carlo computation to
the preliminary measurement by PAMELA at
10~GeV~\cite{Mocchiutti:2009}; the secondary fluxes can be neglected
for this purpose. We find $R_{e^-}^0 = 1.8 \times 10^{50}
\,\text{GeV}^{-1}$ for $\gamma = 2.4$ which corresponds to a total
injection energy of $\int \text{d}E \,E \,R_{e^-}(E) \simeq 7 \times
10^{47} \,\text{erg}$. Solar modulation is treated in the spherical
force--field approximation~\cite{Gleeson:1968zz} with a potential of
$\phi = 600 \, \text{MV}$.

\begin{figure}[t!]\centering
  \psfig{file=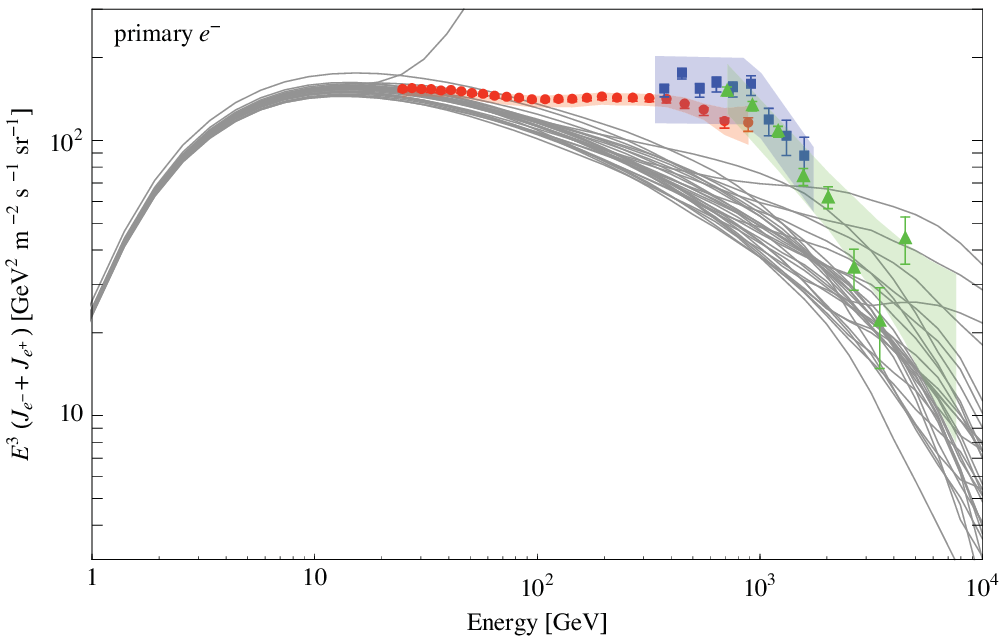,width=0.49\textwidth}
  \psfig{file=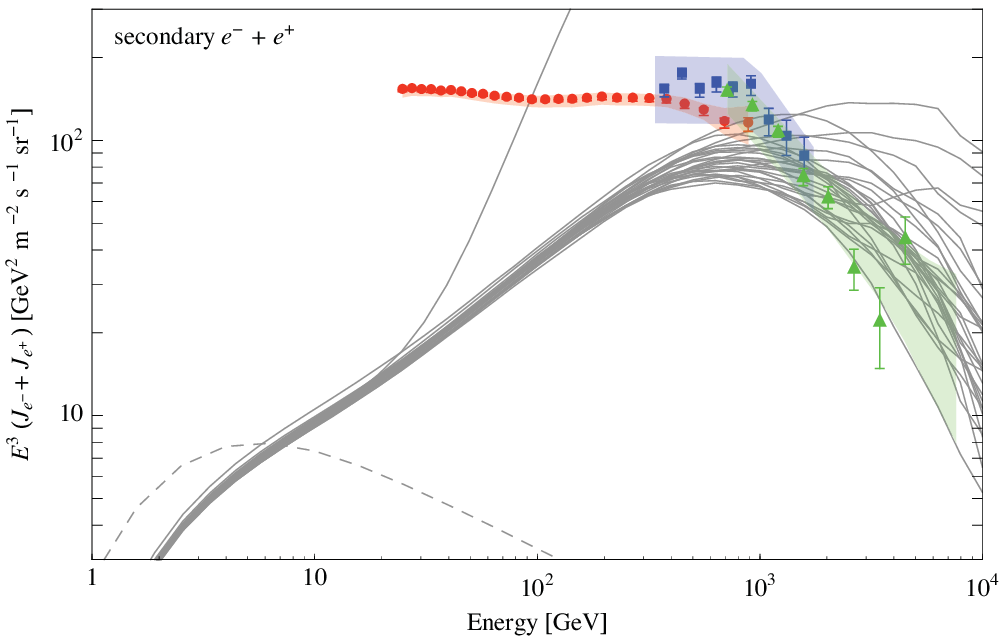,width=0.49\textwidth}
  \caption{Predicted spectra of $e^{-}$ and $e^{+}$ with data from
    Fermi--LAT \cite{Abdo:2009zk} (red circles) and HESS
    \cite{Collaboration:2008aaa, Aharonian:2009ah} (blue squares \&
    green triangles).  {\bf Left:} Primary $e^{-}$ after propagation
    to Earth. {\bf Right:} Secondary $e^{-}$ and $e^{+}$ from cosmic
    ray interactions, created during propagation (dashed line) and
    created during acceleration in SNRs (full lines).}
  \label{mer:E3J}
\end{figure}

The primary electron fluxes on Earth from a large number of
realisations of the source distribution are shown in the left panel of
Fig.~\ref{mer:E3J} and clearly exhibit a deficit with respect to the
measurements by Fermi--LAT and HESS.

\section{Additional positrons from the acceleration of secondaries}
\label{mer:additional}

It is usually assumed that secondary $e^{-}$ and $e^{+}$ are produced
only through spallation (mainly on interstellar hydrogen and helium)
during the galactic propagation of hadronic cosmic rays. At energies
above $10 \, \text{GeV}$, the spectrum of such secondaries after
propagation should be softer than that of the primaries (by the power
$\delta$, the spectral index of the diffusion coefficient), and the
positron fraction should therefore be falling. We calculate secondary
$e^{+}$ from propagation of protons and nuclei following
Ref.~\refcite{Delahaye:2008ua}, using however the propagation
parameters specified above.

It was recently suggested \cite{Blasi:2009hv} that the acceleration of
secondary $e^{-}$ and $e^{+}$ in the cosmic ray sources, i.e. SNRs,
could lead to a harder source spectrum of $e^{+}$ that can explain the
observed rise in the positron fraction. As the volume in which
particles that participate in DSA grows with the diffusion coefficient
$D_{\mathrm{Bohm}}(p) \propto p$, the fraction of the secondaries
which can participate in DSA also increases linearly with
momentum. The resulting source spectrum $R_\pm$ of secondary $e^{-}$
and $e^{+}$ is then a sum of two power laws, corresponding to the
unaccelerated and the accelerated secondaries:
\begin{equation}
  R_\pm \simeq R_\pm^0 \, p^{-\gamma} \left[1 
    + \left(\frac{p}{p_\text{cross}}\right)\right] ,
\end{equation}
where the ``cross-over'' momentum, $p_\text{cross}$, satisfies
\begin{equation}
\label{mer:pxdef}
D (p_\text{cross}) = \frac{3}{4}
\frac{ru_1^2\tau_\text{SNR}}{(\gamma + 2)(1/\xi + r^2)}\,.
\end{equation}
As has been noted \cite{Blasi:2009hv}, this mechanism is most
efficient for {\it old} SNRs where field amplification by the shock
wave is not very effective anymore. We therfore introduce a fudge
factor $K_\text{B}$ that parameterises the effect of the smaller field
amplification on the otherwise Bohm-like diffusion coefficient in the
SNR,
\begin{equation}
  D (E) = 3.3 \times 10^{22} K_\text{B} \, \bigg(\frac{B}{\mu{\rm G}}\bigg)^{-1} 
  \bigg(\frac{E}{\text{GeV}}\bigg) \, \text{cm}^2 \text{s}^{-1}.
\label{mer:D(p)}
\end{equation}

A break in the source spectrum occurs at $p_{\mathrm{break}}$ because
the growth of the acceleration zone is bounded by the physics size of
the SNR. The source spectrum $R_\pm$ thus returns to a $p^{-\gamma}$
dependence around $p = p_\text{break}$. At even higher energies the
secondary spectrum cuts off at the same $E_\text{cut}$ as the primary
$e^{-}$ (see Sec.~\ref{mer:PrimaryElectrons}).

Following Refs.~\refcite{Blasi:2009hv, Blasi:2009bd}, the parameters
are chosen to be: $u_1 = 0.5 \times 10^8 \,\text{cm} \,\text{s}^{-1}$,
$n_{\text{gas}, 1} = 2 \,\text{cm}^{-3}$, $B = 1
\,\mu\text{G}$. Instead of fixing the normalisation of the injection
spectrum {\it ad hoc}, we determine it from $\gamma$--ray observations
of SNRs. Knowing the cross-sections for the production of
$\gamma$-rays (from the decay of neutral pions) and the production of
secondary $e^{-}$ and $e^{+}$ (from the decay of charged pions), we
determine the normalisation
\begin{align}
  R_+^0 = 7.4 \times 10^{48} \bigg(\frac{\tau_\text{SNR}}{10^4
    \text{yr}}\bigg) \bigg(\frac{Q_\gamma^0}{5.7 \times 10^{33}
    \text{s}^{-1}\text{TeV}^{-1}}\bigg) \text{GeV}^{-1} .
\end{align}
where $Q_\gamma^0$ is the observed typical luminosity of SNR in
$\gamma$-rays as determined from the same compilation of data as
above.

The right panel of Fig.~\ref{mer:E3J} shows the expected flux of
secondary $e^{-}$ and $e^{+}$ for $30$ realizations of the possible
distribution of SNRs in our Galaxy. Clearly this component can
potentially match the high energy Fermi--LAT and HESS data. We also
show the secondary $e^{-}$ and $e^{+}$ from the propagation of protons
and nuclei, which is subdominant.

\section{Total electron--positron flux and positron fraction}
\label{mer:total}

\begin{figure}[t!]\centering
  \psfig{file=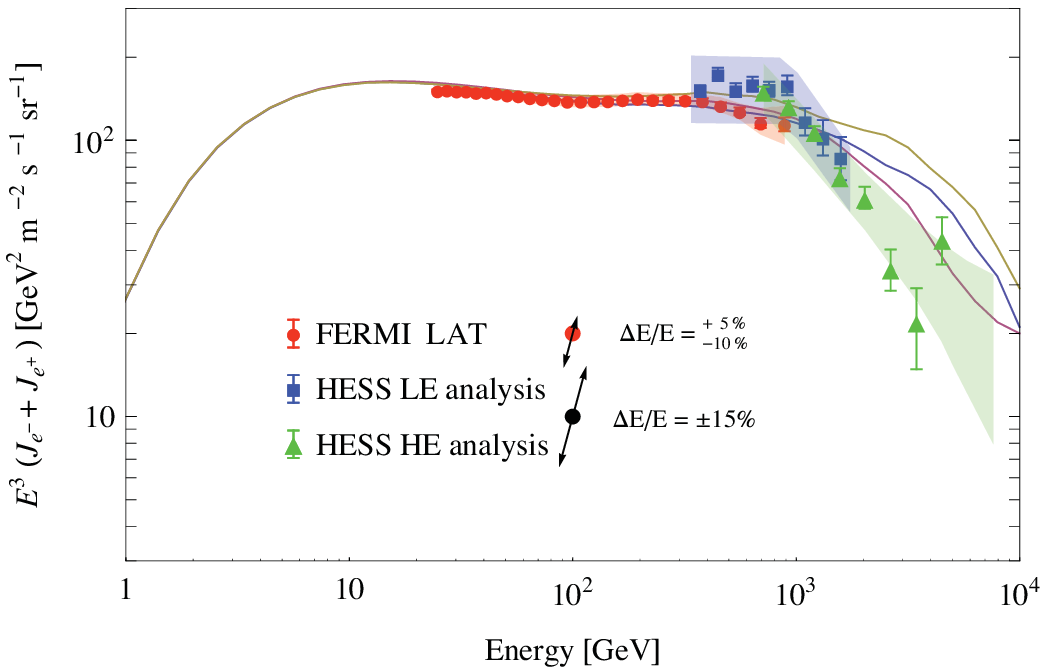,width=0.49\textwidth}
  \psfig{file=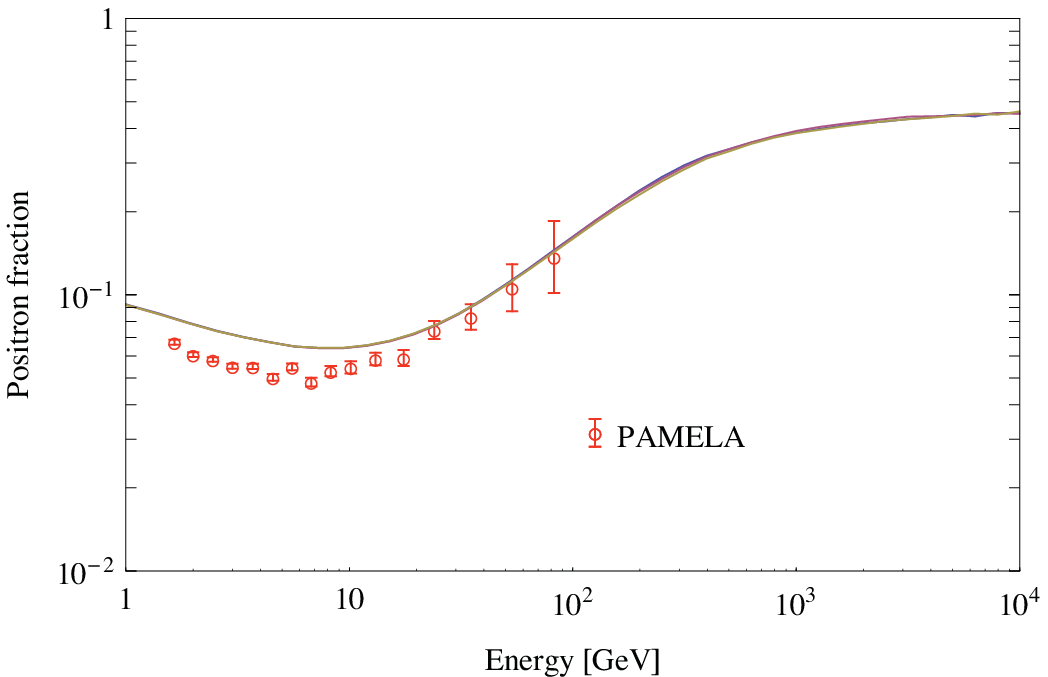,width=0.49\textwidth}
  \caption{{\bf Left:} Predicted spectra of the total flux of primary
    and secondary $e^{-}$ and $e^{+}$.  {\bf Right:} The predicted positron
    fraction after fixing all free parameters using the
    total $(e^{+} + e^{-})$ flux, compared with data from
    PAMELA~\cite{Adriani:2008zr}.}
  \label{mer:E3Jtot}
\end{figure}

Figure~\ref{mer:E3Jtot} shows the total $(e^{+} + e^{-})$ flux obtained
by adding the primary $e^{-}$ and the secondary $e^{-}$ and $e^{+}$
(see. Fig.~\ref{mer:E3J}). As explained above, we selected just those
(3) realisations of the source distribution which give the best fit to
the measurements by Fermi--LAT and HESS. By fitting to the $(e^{+} +
e^{-})$ flux we have also fixed the value for the diffusion
coefficient near the SNR shock wave (which determines the ratio of the
accelerated to the unaccelerated secondaries) to about 15 times the
Bohm value i.e. $K_B \simeq 15$.

Having thus fixed all free parameters we can make an independent
prediction for the positron fraction which is shown in the right panel
of Fig.~\ref{mer:E3Jtot}. It is in very good agreement with the PAMELA
measurements above $\gtrsim 10 \, \text{GeV}$; we do not expect to
match the data below $10 \, \text{GeV}$ because our analytic approach
neglects complications such as convection and reacceleration in the
ISM which can be important at these energies.

\section{Nuclear secondary-to-primary ratios}

The acceleration of secondaries described above modifes not only the
spectra of secondary $e^{-}$ \& $e^{+}$ but also of other charged
secondaries. This provides a direct test of this explanation for the
rise in the positron fraction, e.g. a rise in the antiproton-to-proton
ratio is also predicted~\cite{Blasi:2009bd} (consistent with the
measurements by PAMELA~\cite{Adriani:2008zq} so far). We have argued
\cite{Mertsch:2009ph} that {\em nuclear} secondary-to-primary ratios
can be used to not only test this model but also to discriminate
against alternative explanations such as dark matter
annihilation/decay or nearby pulsars which cannot affect nuclear
abundances.

\begin{figure}[t!]\centering
  \psfig{file=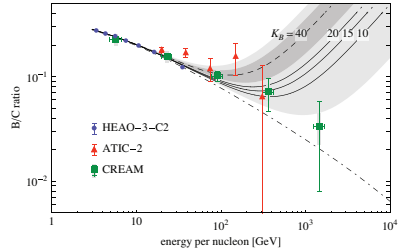,width=0.45\textwidth}\\[0.3cm]
  \caption{The predicted B/C ratio in cosmic rays (from
    Ref.~\refcite{Mertsch:2009ph}) for the `leaky box model' with
    production of secondaries during propagation only (dot-dashed
    line), and including production and acceleration of secondaries in
    nearby SNRs (solid lines) for values of the diffusion coefficient
    near the shock wave which best fit the $e^{\pm}$ spectrum (see
    Sec.~\ref{mer:total}). The dashed line corresponds to the value of
    the diffusion co-efficient required to fit ATIC-2 data on Ti/Fe
    (from Ref.~\refcite{Mertsch:2009ph}), along with its $1 \sigma$
    and $2 \sigma$ error bands. The data points are from HEAO-3-C2
    (circles) \cite{Engelmann:1990zz}, ATIC-2 (triangles)
    \cite{Panov:2007fe} and CREAM (squares) \cite{Ahn:2008my}.}
  \label{mer:B2C}
\end{figure}

We have calculated the rise in nuclear secondary-to-primary ratios
expected from the acceleration of secondaries, in particular, we have
used recent ATIC-2 data on the titanium-to-iron ratio which exhibit a
\emph{rise} around $100 \, \text{GeV}/\mathrm{n}$ to fix the diffusion
coefficient. We find a `fudge factor' of $K_B \simeq 40$ is necessary,
albeit with large error bars due to the limited statistics of the
ATIC-2 data. Again, we can make an independent prediction for a
different secondary-to-primary ratio, namely boron-to-carbon
(B/C). Our prediction, together with the canonical expectation for a
purely secondary origin through spallation in the ISM, are shown in
Fig.~\ref{mer:B2C} with a selection of data. Currently, B/C is being
measured by PAMELA, and it is also a prime goal of the upcoming AMS-02
mission.\cite{AMS02}

\section{Conclusion}

We have discussed the recently measured excesses in the $e^{+}$
fraction and in the total $e^{+} + e^{-}$ flux in galactic cosmic
rays. The fluctuations induced by the discreteness of the cosmic ray
sources has been estimated assuming these to be SNRs. However when the
source spectral index is estimated from $\gamma$-ray observations of
SNRs, the expected total $e^{+} + e^{-}$ flux at Earth is
deficient. Adding the flux of secondary $e^{-}$ and $e^{+}$ produced
and accelerated in nearby SNRs brings the model prediction back in
agreement with the measurements and naturally implies a rise in the
$e^{+}$ fraction. A crucial test of this idea is to determine if
nuclear secondary-to-primary ratios, e.g.  B/C, also increase with
energy. Such nearby SNRs ought also to be detected by the IceCube
neutrino observatory in a few years.\cite{Mertsch:2009ph}. This would
be the first direct astronomical identification of the sources of
galactic cosmic rays.



\end{document}